\documentstyle[preprint,eqsecnum,aps]{revtex}
\begin{document}
\draft
\title{Green's function for the Relativistic Coulomb System via Sum Over
Perturbation Series}
\author{De-Hone Lin \thanks{%
e-mail:d793314@phys.nthu.edu.tw}}
\address{Department of Physics, National Tsing Hua University \\
Hsinchu 30043, Taiwan}
\date{\today}
\maketitle
\begin{abstract}
We evaluate the
Green's function of the D-dimensional relativistic Coulomb system via sum
over perturbation series which is obtained by expanding the exponential
containing the potential term $V({\bf x)}$ in the path integral into a power
series. The energy spectra and wave functions are extracted from the
resulting amplitude. 
\end{abstract}
\newpage \tolerance=10000
\section{Introduction}
~~~~~~Most physical problems cannot be solved exactly. It is therefore
necessary to develop approximation procedures which allow us to approach the
exact result with appropriate accuracy. An important approximation method
for solving problems in quantum mechanics (QM) is the
Rayleigh-Schr\"{o}dinger perturbation theory. It provides us an effective
method to calculate the approximate solutions of many problems which can not
be exactly solved by using the Schr\"{o}dinger equation. Similar to the
standard QM, the perturbation method can be developed in the path integral
framework of QM \cite{1}. Historically of utmost importance was the
application of the perturbation expansion of path integral to the quantum
electrodynamics by Feynman \cite{2}, from which he derived for the first
time the ``Feynman's rules'', which provide an extremely effective method to
calculate the perturbation series and a clear, neat interpretation of the
interaction picture.

In the past 10 years, perturbation expansion of the path integral has been
used to obtain the exact Green's functions for $\delta $-function potential
problems \cite{3,3.5,4,4.5}, non-relativistic Coulomb system \cite{5}, and
to yield the Dirichlet boundary conditions 
in Refs. \cite{6, 6.5} for the non-relativistic problems 
and in Ref. \cite{7} for the relativistic problems 
by summing the $\delta $-function
perturbation series.

In this paper, we would like to add a further application of the
perturbation method of the path integral. We calculate the Green's function
of a D-dimensional relativistic Coulomb system via summing over the
perturbation series. The energy spectra and wave functions are extracted
from the resulting amplitude.

\section{Path integral for the relativistic Coulomb system via sum over the
perturbation series}

~~~~~~Let us first consider a point particle of mass $M$ moving at a
relativistic velocity in a $(D+1)$-dimensional Minkowski space with a given
electromagnetic field. By using $t=-i\tau =-ix^{4}/c$, the path integral
representation of the Green's function is conveniently formulated in a $(D+1)
$-Euclidean spacetime with the Euclidean metric, 
\begin{equation}
(g_{\mu \nu })={\rm diag}\;\,(1,\cdots ,1,c^{2}),  \label{1.1}
\end{equation}
and it is given by \cite{8,9} 
\begin{equation}
G({\bf {x}}_{b},{\bf {x}}_{a};E)=\frac{i\hbar }{2Mc}\int_{0}^{\infty }dS\int
D\rho \Phi \left[ \rho \right] \int D^{D}xe^{-A_{E}/\hbar }.\qquad \qquad 
\label{1.2}
\end{equation}
The action integral 
\begin{equation}
A_{E}=\int_{\lambda _{a}}^{\lambda _{b}}d\lambda \left[ \frac{M}{2\rho
\left( \lambda \right) }{\bf {x}}^{\prime ^{2}}\left( \lambda \right) -i%
\frac{e}{c}{\bf A(x)\cdot x^{\prime }(}\lambda {\bf )}-\rho \left( \lambda
\right) \frac{(E-V({\bf x}))^{2}}{2Mc^{2}}+\rho \left( \lambda \right) \frac{%
Mc^{2}}{2}\right] ,  \label{1.3}
\end{equation}
where $S$ is defined by 
\begin{equation}
S=\int_{\lambda _{a}}^{\lambda _{b}}d\lambda \rho (\lambda ),  \label{1.4}
\end{equation}
in which $\rho (\lambda )$ is an arbitrary dimensionless fluctuating scale
variable, and $\Phi \lbrack \rho ]$ is some convenient gauge-fixing
functional, such as $\Phi \left[ \rho \right] =\delta \left[ \rho -1\right] $%
, to fix the value of $\rho (\lambda )$ to unity \cite{8,9}. $\hbar /Mc$ is
the well-known Compton wave length of a particle of mass $M$, ${\bf A(x)}$
is the vector potential, $V({\bf x})$ is the scalar potential, $E$ is the
system energy, and ${\bf {x}}$ is the spatial part of the ($D+1$) vector $x=(%
{\bf {x}},\tau )$. This path integral forms the basis for studying
relativistic potential problems.

Expanding the potential term $V({\bf x})$ into a power series and
interchanging the order of integration and summation, we obtain the result 
\begin{equation}
G({\bf {x}}_{b},{\bf {x}}_{a};E)=\frac{i\hbar }{2Mc}\int_{0}^{\infty }dS\int
D\rho \Phi \left[ \rho \right] e^{-\frac{1}{\hbar }\int_{\lambda
_{a}}^{\lambda _{b}}d\lambda \rho (\lambda ){\cal E}}K({\bf {x}}_{b},{\bf {x}%
}_{a};\lambda _{b}-\lambda _{a})  \label{1.5}
\end{equation}
with the series expansion of the pseudotime propagator 
\[
K({\bf {x}}_{b},{\bf {x}}_{a};\lambda _{b}-\lambda _{a})=\left\{ 
\vbox to
24pt{}K_{0}+\sum_{n=1}^{\infty }\frac{1}{n!}\left( -\frac{\beta }{\hbar }%
\right) ^{n}\right. 
\]
\[
\times \int D^{D}xe^{-\frac{1}{\hbar }\int_{\lambda _{a}}^{\lambda
_{b}}d\lambda \left[ \frac{M}{2\rho \left( \lambda \right) }{\bf {x}}%
^{\prime ^{2}}\left( \lambda \right) -i\frac{e}{c}{\bf A(x)\cdot x^{\prime }(%
}\lambda {\bf )}-\rho \left( \lambda \right) \frac{V({\bf x})^{2}}{2Mc^{2}}%
\right] } 
\]
\begin{equation}
\times \left. \int_{\lambda _{a}}^{\lambda _{b}}d\lambda _{1}\rho {\bf (}%
\lambda _{1})V({\bf x(}\lambda _{1}))\int_{\lambda _{a}}^{\lambda
_{b}}d\lambda _{2}\rho {\bf (}\lambda _{2})V({\bf x(}\lambda _{2}))\cdots
\int_{\lambda _{a}}^{\lambda _{b}}d\lambda _{n}\rho {\bf (}\lambda _{n})V(%
{\bf x(}\lambda _{n}))\vbox to 24pt{}\right\} ,  \label{1.6}
\end{equation}
where we have defined the quantities $\beta =E/Mc^{2},$ ${\cal E}=($ $%
M^{2}c^{4}-E^{2})/2Mc^{2}$, and 
\begin{equation}
K_{0}({\bf {x}}_{b},{\bf {x}}_{a};\lambda _{b}-\lambda _{a})=\int D^{D}xe^{-%
\frac{1}{\hbar }\int_{\lambda _{a}}^{\lambda _{b}}d\lambda \left[ \frac{M}{%
2\rho \left( \lambda \right) }{\bf {x}}^{\prime ^{2}}\left( \lambda \right)
-i\frac{e}{c}{\bf A(x)\cdot x^{\prime }(}\lambda {\bf )}-\rho \left( \lambda
\right) \frac{V({\bf x})^{2}}{2Mc^{2}}\right] }.  \label{1.7}
\end{equation}
Ordering the $\lambda $ as $\lambda _{1}<\lambda _{2}<\cdots <\lambda
_{n}<\lambda _{b}$ and denoting ${\bf x(}\lambda _{k})={\bf x}_{k},$ the
perturbative series in Eq. (\ref{1.6}) turns into \cite{1} 
\[
K({\bf {x}}_{b},{\bf {x}}_{a};\lambda _{b}-\lambda _{a})=K_{0}({\bf {x}}_{b},%
{\bf {x}}_{a};\lambda _{b}-\lambda _{a})+\sum_{n=1}^{\infty }\left( -\frac{%
\beta }{\hbar }\right) ^{n}\int_{\lambda _{a}}^{\lambda _{b}}d\lambda
_{n}\int_{\lambda _{a}}^{\lambda _{n}}d\lambda _{n-1}\cdots \int_{\lambda
_{a}}^{\lambda _{2}}d\lambda _{1} 
\]
\begin{equation}
\times \int \left[ \prod_{j=0}^{n}K_{0}({\bf {x}}_{j+1},{\bf {x}}%
_{j};\lambda _{j+1}-\lambda _{j})\right] \prod_{k=1}^{n}\rho _{k}V({\bf x}%
_{k})d{\bf x}_{k},  \label{1.8}
\end{equation}
where $\lambda _{0}=\lambda _{a},\lambda _{n+1}=\lambda _{b},{\bf x}_{n+1}=%
{\bf x}_{b},$ and ${\bf x}_{0}={\bf x}_{a}.$ In the case of an attractive
Coulomb potential, we have 
\begin{equation}
{\bf A(x)=}0,\quad V(r)=-\frac{e^{2}}{r}.  \label{1.9}
\end{equation}
The perturbative expansion in Eq. (\ref{1.8}) becomes 
\[
K({\bf {x}}_{b},{\bf {x}}_{a};\lambda _{b}-\lambda _{a})=K_{0}({\bf {x}}_{b},%
{\bf {x}}_{a};\lambda _{b}-\lambda _{a})+\sum_{n=1}^{\infty }\left( \frac{%
\beta e^{2}}{\hbar }\right) ^{n}\int_{\lambda _{a}}^{\lambda _{b}}d\lambda
_{n}\int_{\lambda _{a}}^{\lambda _{n}}d\lambda _{n-1}\cdots \int_{\lambda
_{a}}^{\lambda _{2}}d\lambda _{1} 
\]
\begin{equation}
\times \int \left[ \prod_{j=0}^{n}K_{0}({\bf {x}}_{j+1},{\bf {x}}%
_{j};\lambda _{j+1}-\lambda _{j})\right] \prod_{k=1}^{n}\rho _{k}\frac{d{\bf %
x}_{k}}{r_{k}}.  \label{1.10}
\end{equation}
The corresponding amplitude $K_{0}$ takes the form 
\begin{equation}
K_{0}({\bf {x}}_{b},{\bf {x}}_{a};\lambda _{b}-\lambda _{a})=\int D^{D}xe^{-%
\frac{1}{\hbar }\int_{\lambda _{a}}^{\lambda _{b}}d\lambda \left[ \frac{M}{%
2\rho \left( \lambda \right) }{\bf {x}}^{\prime ^{2}}\left( \lambda \right)
-\rho \left( \lambda \right) \frac{\hbar ^{2}}{2M}\frac{\alpha ^{2}}{r^{2}}%
\right] },  \label{1.11}
\end{equation}
where $\alpha =e^{2}/\hbar c$ is the fine structure constant. We now choose $%
\Phi \left[ \rho \right] =\delta \left[ \rho -1\right] $ to fix the value of 
$\rho (\lambda )$ to unity. The Green's function in Eq. (\ref{1.5}) becomes 
\[
G({\bf {x}}_{b},{\bf {x}}_{a};E)=\frac{i\hbar }{2Mc}\int_{0}^{\infty }dSe^{-%
\frac{{\cal E}}{\hbar }S}\left\{ \vbox to 24pt{}K_{0}({\bf {x}}_{b},{\bf {x}}%
_{a};S)\right. 
\]
\begin{equation}
+\left. \sum_{n=1}^{\infty }\left( \frac{\beta e^{2}}{\hbar }\right)
^{n}\int_{\lambda _{a}}^{\lambda _{b}}d\lambda _{n}\int_{\lambda
_{a}}^{\lambda _{n}}d\lambda _{n-1}\cdots \int_{\lambda _{a}}^{\lambda
_{2}}d\lambda _{1}\int \left[ \prod_{j=0}^{n}K_{0}({\bf {x}}_{j+1},{\bf {x}}%
_{j};\lambda _{j+1}-\lambda _{j})\right] \prod_{k=1}^{n}\frac{d{\bf x}_{k}}{%
r_{k}}\vbox to 24pt{}\right\} .  \label{1.12}
\end{equation}
We observe that the integration over $S$ is a Laplace transformation.
Because of the convolution property of the Laplace transformation, we obtain 
\[
G({\bf {x}}_{b},{\bf {x}}_{a};E)=\frac{i\hbar }{2Mc} 
\]
\begin{equation}
\times \left\{ G_{0}({\bf {x}}_{b},{\bf {x}}_{a};{\cal E})+\sum_{n=1}^{%
\infty }\left( \frac{\beta e^{2}}{\hbar }\right) ^{n}\int \left[
\prod_{j=0}^{n}G_{0}({\bf {x}}_{j+1},{\bf {x}}_{j};{\cal E})\right]
\prod_{k=1}^{n}\frac{d{\bf x}_{k}}{r_{k}}\right\} .  \label{1.13}
\end{equation}
We now perform the angular decomposition of Eq. (\ref{1.13}) \cite{9,10,11}.
This can be reached by inserting in Eq. (\ref{1.13}) the expansion of $G_{0}$
in term of the D-dimensional hyperspherical harmonics $Y_{l{\bf m}}({\bf 
\hat{x}})$ \cite{12}: 
\begin{equation}
G_{0}({\bf {x}}_{j+1},{\bf {x}}_{j};{\cal E})=\frac{M}{\hbar
(r_{j+1}r_{j})^{D/2-1}}\sum_{l=0}^{\infty }g_{l}^{0}(r_{j+1},r_{j};{\cal E}%
)\sum_{{\bf m}}Y_{l{\bf m}}({\bf \hat{x}}_{j+1})Y_{l{\bf m}}^{\ast }({\bf 
\hat{x}}_{j}),  \label{1.14}
\end{equation}
where the $g_{l}^{0}$ is given by \cite{11} 
\begin{equation}
\int_{0}^{\infty }\frac{dS}{S}e^{-\frac{{\cal E}}{\hbar }%
S}e^{-M(r_{j+1}^{2}+r_{j}^{2})/2\hbar S}I_{\sqrt{(l+D/2-1)^{2}-\alpha ^{2}}%
}\left( \frac{M}{\hbar }\frac{r_{j+1}r_{j}}{S}\right) .  \label{1.15}
\end{equation}
The notation $I$ denotes the modified Bessel function. Integrating over the
intermediate angular part of Eq. (\ref{1.13}), we arrive at 
\begin{equation}
G({\bf {x}}_{b},{\bf {x}}_{a};E)=\frac{i\hbar }{2Mc}\sum_{l=0}^{\infty
}G_{l}(r_{b},r_{a};{\cal E})\sum_{{\bf m}}Y_{l{\bf m}}({\bf \hat{x}}_{b})Y_{l%
{\bf m}}^{\ast }({\bf \hat{x}}_{a}).  \label{1.16}
\end{equation}
The pure radial amplitude $G_{l}(r_{b},r_{a};{\cal E})$ has the form 
\begin{equation}
G_{l}(r_{b},r_{a};{\cal E})=\frac{M}{\hbar }\frac{1}{(r_{b}r_{a})^{D/2-1}}%
\sum_{n=0}^{\infty }\left( \frac{M\beta e^{2}}{\hbar ^{2}}\right)
^{n}g_{l}^{(n)}(r_{b},r_{a};{\cal E})  \label{1.17}
\end{equation}
with $g_{l}^{(n)}$ is given by 
\begin{equation}
g_{l}^{(n)}(r_{b},r_{a};{\cal E})=\int_{0}^{\infty }\cdots \int_{0}^{\infty }%
\left[ \prod_{j=0}^{n}g_{l}^{(0)}(r_{j+1},r_{j};{\cal E})\right]
\prod_{k=1}^{n}dr_{k}.  \label{17.5}
\end{equation}
To obtain the explicit result of $g_{l}^{(n)}$, we note that 
\[
\int_{0}^{\infty }\frac{dS}{S}e^{-\frac{{\cal E}}{\hbar }%
S}e^{-M(r_{b}^{2}+r_{a}^{2})/2\hbar S}I_{\sqrt{(l+D/2-1)^{2}-\alpha ^{2}}%
}\left( \frac{M}{\hbar }\frac{r_{b}r_{a}}{S}\right) 
\]
\begin{equation}
=2\int_{0}^{\infty }dz\frac{1}{\sinh z}e^{-\kappa (r_{b}+r_{a})\coth z}I_{2%
\sqrt{(l+D/2-1)^{2}-\alpha ^{2}}}\left( \frac{2\kappa \sqrt{r_{b}r_{a}}}{%
\sinh z}\right)  \label{1.18}
\end{equation}
with $\kappa =\sqrt{M^{2}c^{4}-E^{2}}/\hbar c$. The equality in Eq. (\ref
{1.18}) can be easily proved by the formulas 
\[
\int_{0}^{\infty }dy\frac{e^{2\nu y}}{\sinh y}\exp \left[ -\frac{t}{2}\left(
\zeta _{a}+\zeta _{b}\right) \coth y\right] I_{\mu }\left( \frac{t\sqrt{%
\zeta _{b}\zeta _{a}}}{\sinh y}\right) 
\]
\begin{equation}
=\frac{\Gamma \left( \left( 1+\mu \right) /2-\nu \right) }{t\sqrt{\zeta
_{b}\zeta _{a}}\Gamma \left( \mu +1\right) }W_{\nu ,\mu /2}\left( t\zeta
_{b}\right) M_{\nu ,\mu /2}\left( t\zeta _{a}\right) ,  \label{1.20}
\end{equation}
with the range of validity 
\[
\begin{array}{l}
\zeta _{b}>\zeta _{a}>0, \\ 
{Re}[(1+\mu )/2-\nu ]>0, \\ 
{Re}(t)>0,\mid \arg t\mid <\pi ,
\end{array}
\]
where $M_{\mu ,\nu }$ and $W_{\mu ,\nu }$ are the Whittaker functions, and 
\begin{equation}
\int_{0}^{\infty }\frac{dy}{y}e^{-zy}e^{-(a^{2}+b^{2})/y}I_{\nu }(\frac{2ab}{%
y})=2I_{\nu }(2a\sqrt{z})K_{\nu }(2b\sqrt{z}),  \label{1.21}
\end{equation}
with the range of validity 
\[
a<b,\quad {Re}z>0. 
\]
From Eq. (\ref{1.18}), we get, by using the formula 
\begin{equation}
\int_{0}^{\infty }drre^{-r^{2}/a}I_{\nu }(\varsigma r)I_{\nu }(\xi r)=\frac{a%
}{2}e^{a(\xi ^{2}+\varsigma ^{2})/4}I_{\nu }\left( \frac{a\xi \varsigma }{2}%
\right) ,  \label{21.5}
\end{equation}
the result 
\[
g_{l}^{(1)}(r_{b},r_{a};{\cal E})=\int_{0}^{\infty }g_{l}^{(0)}(r_{b},r;%
{\cal E})g_{l}^{(0)}(r,r_{a};{\cal E})dr 
\]
\begin{equation}
=\frac{2^{2}}{\kappa }\int_{0}^{\infty }zh(z)dz,  \label{1.22}
\end{equation}
where the function $h(z)$ is defined as 
\begin{equation}
h(z)=\frac{1}{\sinh z}e^{-\kappa (r_{b}+r_{a})\coth z}I_{2\sqrt{%
(l+D/2-1)^{2}-\alpha ^{2}}}\left( \frac{2\kappa \sqrt{r_{b}r_{a}}}{\sinh z}%
\right) .  \label{1.23}
\end{equation}
The expression for $g_{l}^{(n)}(r_{b},r_{a};{\cal E})$ can be obtained by
induction with respect to $n$, and is given by 
\begin{equation}
g_{l}^{(n)}(r_{b},r_{a};{\cal E})=\frac{2^{n+1}}{n!}\frac{1}{\kappa ^{n}}%
\int_{0}^{\infty }z^{n}h(z)dz.  \label{1.25}
\end{equation}
Inserting the expression in Eq. (\ref{1.17}), we obtain 
\[
G_{l}(r_{b},r_{a};{\cal E})=\frac{M}{\hbar }\frac{2}{(r_{b}r_{a})^{D/2-1}} 
\]
\begin{equation}
\times \int_{0}^{\infty }dze^{\left( \frac{2M\beta e^{2}}{\hbar ^{2}\kappa }%
\right) z}\frac{1}{\sinh z}e^{-\kappa (r_{b}+r_{a})\coth z}I_{2\sqrt{%
(l+D/2-1)^{2}-\alpha ^{2}}}\left( \frac{2\kappa \sqrt{r_{b}r_{a}}}{\sinh z}%
\right) .  \label{1.26}
\end{equation}
With help of the formula in Eq. (\ref{1.20}), we complete the integration of
Eq. (\ref{1.26}), and find the radial Green's function for $r_{b}>r_{a}$ in
the closed form, 
\[
G_{l}(r_{b},r_{a};E)=\frac{1}{(r_{b}r_{a})^{(D-1)/2}}\frac{Mc}{\sqrt{%
M^{2}c^{4}-E^{2}}}\qquad \qquad \qquad 
\]
\[
\times \frac{\Gamma \left( 1/2+\sqrt{(l+D/2-1)^{2}-\alpha ^{2}}-\frac{%
E\alpha }{\sqrt{M^{2}c^{4}-E^{2}}}\right) }{\Gamma \left( 1+2\sqrt{%
(l+D/2-1)^{2}-\alpha ^{2}}\right) } 
\]
\[
\times W_{\frac{E\alpha }{\sqrt{M^{2}c^{4}-E^{2}}},\sqrt{(l+D/2-1)^{2}-%
\alpha ^{2}}}\left( \frac{2}{\hbar c}\sqrt{M^{2}c^{4}-E^{2}}r_{b}\right) 
\]
\begin{equation}
\times M_{\frac{E\alpha }{\sqrt{M^{2}c^{4}-E^{2}}},\sqrt{(l+D/2-1)^{2}-%
\alpha ^{2}}}\left( \frac{2}{\hbar c}\sqrt{M^{2}c^{4}-E^{2}}r_{a}\right)
\label{1.27}
\end{equation}

The energy spectra and wave functions can be extracted from the poles of Eq.
(\ref{1.27}). For convenience, we define the following variables 
\begin{equation}
\left\{ 
\begin{array}{l}
\begin{array}{rcl}
\kappa & = & \frac{1}{\hbar c}\sqrt{M^{2}c^{4}-E^{2}},
\end{array}
\\ 
\begin{array}{rcl}
\nu & = & \frac{\alpha E}{\sqrt{M^{2}c^{4}-E^{2}}},
\end{array}
\\ 
\begin{array}{rcl}
\tilde{l} & = & \sqrt{(l+D/2-1)^{2}-\alpha ^{2}}-1/2,
\end{array}
\end{array}
\right.  \label{1.28}
\end{equation}
From the poles of $G_{l}(r_{b},r_{a};E)$, we find that the energy levels
must satisfy the equality 
\begin{equation}
-\nu +\tilde{l}+1=-n_{r},\quad n_{r}=0,1,2,3,\cdots .  \label{1.29}
\end{equation}
Expanding this equation into powers of $\alpha $, we get 
\[
E_{nl}\approx \pm Mc^{2}\left\{ 1-\frac{1}{2}\left[ \frac{\alpha }{n+1/2(D-3)%
}\right] ^{2}-\frac{\alpha ^{4}}{\left[ n+1/2(D-3)\right] ^{3}}\right. 
\]
\begin{equation}
\left. \times \left[ \frac{1}{2\left[ l+1/2(D-2)\right] }-\frac{3}{8}\frac{1%
}{\left[ n+1/2(D-3)\right] }\right] +O(\alpha ^{6})\right\} .  \label{1.30}
\end{equation}
Here $n$ is defined by $n_{r}=n-l-1$. We point out that by setting D=3, the
energy levels reduce to the well-known form 
\begin{equation}
E_{nl}\approx \pm Mc^{2}\left\{ 1-\frac{1}{2}\left( \frac{\alpha }{n}\right)
^{2}-\frac{\alpha ^{4}}{n^{3}}\left[ \frac{1}{2l+1}-\frac{3}{8n}\right]
+O(\alpha ^{6})\right\} .  \label{1.31}
\end{equation}

The pole positions, which satisfy $\nu =\tilde{n}_{l}\equiv n+\tilde{l}-l$ ($%
n=l+1,l+2,l+3,\cdots $), correspond to the bound states of the D-dimensional
relativistic Coulomb system. Near the positive-energy poles, we use the
behavior for $\nu \approx \tilde{n}_{l}$, 
\begin{equation}
-\Gamma (-\nu +\tilde{l}+1)\frac{M}{\hbar \kappa }\approx \frac{(-)^{n_{r}}}{%
\tilde{n}_{l}^{2}n_{r}!}\frac{1}{\tilde{a}_{H}}\left( \frac{E}{Mc^{2}}%
\right) ^{2}\frac{2\hbar Mc^{2}}{E^{2}-E_{nl}^{2}}  \label{1.32}
\end{equation}
with $\tilde{a}_{H}\equiv a_{H}\frac{Mc^{2}}{E}$ being the modified
energy-dependent Bohr radius and $n_{r}=n-l-1$ the radial quantum number, to
extract the wave functions of the D-dimensional Coulomb system 
\[
G_{l}(r_{b},r_{a};E)=-\frac{i}{(r_{b}r_{a})^{(D-1)/2}}\sum_{n=l+1}^{\infty
}\left( \frac{E}{Mc^{2}}\right) ^{2}\frac{2\hbar Mc^{2}}{E^{2}-E_{nl}^{2}}%
\qquad \qquad 
\]
\[
\times \frac{1}{\left[ (2\tilde{l}+1)!\right] ^{2}}\frac{1}{\tilde{n}_{l}^{2}%
\tilde{a}_{H}}\frac{(\tilde{n}_{l}+\tilde{l})!}{(n-l-1)!}e^{-(r_{b}+r_{a})/%
\tilde{a}_{H}\tilde{n}_{l}}\left( \frac{2r_{b}}{\tilde{a}_{H}\tilde{n}_{l}}%
\frac{2r_{a}}{\tilde{a}_{H}\tilde{n}_{l}}\right) ^{\tilde{l}+1} 
\]
\[
\times M(-n+l+1,2\tilde{l}+2;\frac{2r_{b}}{\tilde{a}_{H}\tilde{n}_{l}}%
)M(-n+l+1,2\tilde{l}+2;\frac{2r_{a}}{\tilde{a}_{H}\tilde{n}_{l}})\label{1.33}
\]
\begin{equation}
=-\frac{i}{(r_{b}r_{a})^{(D-1)/2}}\sum_{n=l+1}^{\infty }(\frac{E}{Mc^{2}}%
)^{2}\frac{2\hbar Mc^{2}}{E^{2}-E_{nl}^{2}}R_{nl}(r_{b})R_{nl}^{*}(r_{a})+%
\cdots ,  \label{1.34}
\end{equation}
where we have expressed the Whittaker function $M_{\lambda ,\mu }(z)$ in
terms of the Kummer functions $M(a,b;z)$, 
\begin{equation}
M_{\lambda ,\mu }(z)=z^{\mu +1/2}e^{-z/2}M(\mu -\lambda +1/2,2\mu +1;z).
\label{1.35}
\end{equation}
From this we obtain the radial wave functions 
\[
R_{nl}(r)=\frac{1}{\tilde{n}_{l}\tilde{a}_{H}^{1/2}}\frac{1}{(2\tilde{l}+1)!}%
\sqrt{\frac{(\tilde{n}_{l}+\tilde{l})!}{(n-l-1)!}}\qquad \qquad \qquad
\qquad 
\]
\begin{equation}
\times \left( \frac{2r}{\tilde{a}_{H}\tilde{n}_{l}}\right) ^{^{\tilde{l}%
+1}}e^{-r/\tilde{a}_{H}\tilde{n}_{l}}M(-n+l+1,2\tilde{l}+2;\frac{2r}{\tilde{a%
}_{H}\tilde{n}_{l}}).\qquad \qquad  \label{1.36}
\end{equation}
The normalized wave functions are given by 
\begin{equation}
\Psi _{nl{\bf m}}({\bf x})=\frac{1}{r^{(D-1)/2}}R_{nl}(r)Y_{l{\bf m}}({\bf 
\hat{x}}).  \label{1.37}
\end{equation}

Before extracting the continuous wave function we note that the parameter $%
\kappa $ is real for $\mid E\mid <Mc^{2}$. For $\mid E\mid >Mc^{2}$, the
square root in Eq. (\ref{1.28}) has two imaginary solutions 
\begin{equation}
\kappa =\mp i\tilde{k},\quad \tilde{k}=\frac{1}{\hbar c}\sqrt{%
E^{2}-M^{2}c^{4}},  \label{1.38}
\end{equation}
corresponding to 
\begin{equation}
\nu =\pm i\tilde{\nu},\quad \tilde{\nu}=\frac{E\alpha }{\hbar c\tilde{k}}%
.\quad \qquad \qquad  \label{1.39}
\end{equation}
Therefore the amplitude has a right-handed cut for $E>Mc^{2}$ and $E<-Mc^{2}$%
. For simplicity, we will only consider the positive energy cut.

The continuous wave function are recovered from the discontinuity of the
amplitudes $G_{l}(r_{b},r_{a};E)$ across the cut in the complex $E$ plane.
Hence we have 
\[
{\rm disc}G_{l}(r_{b},r_{a};E>Mc^{2})=G_{l}(r_{b},r_{a};E+i\eta
)-G_{l}(r_{b},r_{a};E-i\eta )=-\frac{i}{(r_{b}r_{a})^{(D-1)/2}} 
\]
\begin{equation}
\times \frac{M}{\hbar \tilde{k}}\left[ \frac{\Gamma (-i\tilde{\nu}+\tilde{l}%
+1)}{(2\tilde{l}+1)!}W_{i\tilde{\nu},\tilde{l}+1/2}(-2i\tilde{k}r_{b})M_{i%
\tilde{\nu},\tilde{l}+1/2}(-2i\tilde{k}r_{a})+(\tilde{\nu}\rightarrow -%
\tilde{\nu})\right] .  \label{1.40}
\end{equation}
Using the relations 
\begin{equation}
M_{\kappa ,\mu }(z)=e^{\pm i\pi (2\mu +1)/2}M_{-\kappa ,\mu }(-z),
\label{1.41}
\end{equation}
where the sign is positive or negative depending on whether Im$z>0$ or Im$%
z<0 $, and 
\[
W_{\lambda ,\mu }(z)=e^{i\pi \lambda }e^{-i\pi (\mu +1/2)}\frac{\Gamma (\mu
+\lambda +1/2)}{\Gamma (2\mu +1)} 
\]
\begin{equation}
\times \left[ M_{\lambda ,\mu }(z)-\frac{\Gamma (2\mu +1)}{\Gamma (\mu
-\lambda +1/2)}e^{-i\pi \lambda }W_{-\lambda ,\mu }(e^{-i\pi }z)\right] ,
\label{1.42}
\end{equation}
which is valid only for arg $(z)\in (-\pi /2,3\pi /2)$ and $\quad 2\mu \neq
-1,-2,-3,\cdots $. The discontinuity of the amplitude is found to be 
\[
{\rm disc}G_{l}(r_{b},r_{a};E>Mc^{2})=-\frac{i}{(r_{b}r_{a})^{(D-1)/2}}\frac{%
M}{\hbar \tilde{k}}\frac{\mid \Gamma (-i\tilde{\nu}+\tilde{l}+1)\mid ^{2}}{%
\mid \Gamma (2\tilde{l}+2)\mid ^{2}} 
\]
\begin{equation}
\times e^{\pi \tilde{\nu}}M_{-i\tilde{\nu},\tilde{l}+1/2}(2i\tilde{k}%
r_{b})M_{i\tilde{\nu},\tilde{l}+1/2}(-2i\tilde{k}r_{a}).  \label{1.43}
\end{equation}
Thus we have 
\[
\int_{Mc^{2}}^{\infty }\frac{dE}{2\pi \hbar }{\rm disc}%
G_{l}(r_{b},r_{a};E>Mc^{2})\qquad \qquad \qquad \qquad \qquad 
\]
\[
=\frac{1}{2\pi \hbar }\int_{-\infty }^{\infty }\frac{(\hbar c)^{2}\tilde{k}d%
\tilde{k}}{\sqrt{M^{2}c^{4}+(\hbar c\tilde{k})^{2}}}{\rm disc}%
G_{l}(r_{b},r_{a};E>Mc^{2}) 
\]
\begin{equation}
=-\frac{i}{(r_{b}r_{a})^{(D-1)/2}}\int_{-\infty }^{\infty }d\tilde{k}(\frac{E%
}{Mc^{2}})R_{\tilde{k}l}(r_{b})R_{\tilde{k}l}^{*}(r_{a}).\qquad \qquad
\label{1.44}
\end{equation}
From this, we obtain the continuous radial wave function of the
D-dimensional relativistic Coulomb system 
\begin{equation}
R_{\tilde{k}l}(r)=\sqrt{\frac{1}{2\pi }}\frac{1}{\left[ 1+(\frac{c\hbar 
\tilde{k}}{Mc^{2}})^{2}\right] ^{1/2}}\frac{\mid \Gamma (-i\tilde{\nu}+%
\tilde{l}+1)\mid }{(2\tilde{l}+1)!}e^{\pi \tilde{\nu}/2}M_{i\tilde{\nu},%
\tilde{l}+1/2}(-2i\tilde{k}r)
\end{equation}
\[
=\sqrt{\frac{1}{2\pi }}\frac{1}{\left[ 1+(\frac{c\hbar \tilde{k}}{Mc^{2}}%
)^{2}\right] ^{1/2}}\frac{\mid \Gamma (-i\tilde{\nu}+\tilde{l}+1)\mid }{(2%
\tilde{l}+1)!}\qquad \qquad \qquad \qquad 
\]
\begin{equation}
\times e^{\pi \tilde{\nu}/2}e^{i\tilde{k}r}(-2i\tilde{k}r)^{\tilde{l}%
+1}\times M(-i\tilde{\nu}+\tilde{l}+1,2\tilde{l}+2;-2i\tilde{k}r).\qquad
\qquad \qquad  \label{1.45}
\end{equation}
It is easy to check the result is in accordance with the non-relativistic
wave function when we take the non-relativistic limit.

\section{Concluding remarks}

~~~~~~~In this paper we have calculated the Green's function of the
relativistic Coulomb system via sum over perturbation series. From the
resulting amplitude, the energy levels and wave functions are given.
Different from the conventional treatment in path integral using the
space-time and Kustaanheimo-Stiefel transformation techniques (e.g. \cite
{9,11}), the method presented here just involves the computation of the
expectation value of the moments $Q^{n}$ $(Q=\int_{\lambda _{a}}^{\lambda
_{b}}d\lambda \rho {\bf (}\lambda )V({\bf x(}\lambda )))$ over the measure 
\[
K_{0}({\bf {x}}_{b},{\bf {x}}_{a};\lambda _{b}-\lambda _{a})=\int D^{D}xe^{-%
\frac{1}{\hbar }\int_{\lambda _{a}}^{\lambda _{b}}d\lambda \left[ \frac{M}{%
2\rho \left( \lambda \right) }{\bf {x}}^{\prime ^{2}}\left( \lambda \right)
-\rho \left( \lambda \right) \frac{V({\bf x})^{2}}{2Mc^{2}}\right] } 
\]
and summing them in accordance with the Feynman-Kac formula \cite{14} 
\[
G({\bf {x}}_{b},{\bf {x}}_{a};E)=\frac{i\hbar }{2Mc}\int_{0}^{\infty }dS\int
D\rho \Phi \left[ \rho \right] e^{-\frac{1}{\hbar }\int_{\lambda
_{a}}^{\lambda _{b}}d\lambda \rho (\lambda ){\cal E}} 
\]
\begin{equation}
\times E\left[ \exp \left\{ -\frac{1}{\hbar }\int_{\lambda _{a}}^{\lambda
_{b}}d\lambda \rho {\bf (}\lambda )V({\bf x(}\lambda ))\right\} \right]
\end{equation}
\[
=\frac{i\hbar }{2Mc}\int_{0}^{\infty }dS\int D\rho \Phi \left[ \rho \right]
e^{-\frac{1}{\hbar }\int_{\lambda _{a}}^{\lambda _{b}}d\lambda \rho (\lambda
){\cal E}} 
\]
\begin{equation}
\times \sum_{n=1}^{\infty }\frac{(-\beta /\hbar )}{n!}^{n}E\left[ \left(
\int_{\lambda _{a}}^{\lambda _{b}}d\lambda \rho {\bf (}\lambda )V({\bf x(}%
\lambda ))\right) ^{n}\right] ,  \label{1.46}
\end{equation}
where the notation $E\left[ \star \right] $ stands for the expectation value
of the moment $\star $.

We hope that the procedure presented in this article may help us to obtain
the results of other interesting relativistic systems. \newline
\centerline{ACKNOWLEDGMENTS} \newline
{The author is grateful to Doctor M. C. Chang for helpful discussions. The
work is supported by the National Youth Council of the ROC under contract
number NYC300375.}


\newpage

\begin{thebibliography}{99}
\bibitem{1}  R. P. Feynman, and A. Hibbs, ``{\it Quantum Mechanics and Path
Integrals}'', McGraw Hill, New Yark, 1965.

\bibitem{2}  R. P. Feynman, Phys. Rev. {\bf 84}, 108 (1951).

\bibitem{3}  D. Bauch, Nuovo Cim. {\bf B {85}}, 118 (1985).

\bibitem{3.5}  B. Gaveau, and L. S. Schulman, J. Phys. {\bf A {19}}, 1833
(1986), different from the perturbation treatment, the propagator of the $%
\delta $-function potential in this paper is given by the functional
integral approach based on Feynman-Kac formula.

\bibitem{4}  S. V. Lawande, and K. V. Bhagwat, Phys. Lett. {\bf A {131}}, 8
(1988).

\bibitem{5}  D. C. Khandekar, S. V. Lawande, and K. V. Bhagwat, ``{\it %
PATH-INTEGRAL METHODS and their APPLICATIONS}'', World Scientific,
Singapore, 1993.

\bibitem{4.5}  C. Grosche, J. Phys. {\bf A 23}, 5205 (1990).

\bibitem{6}  T. E. Clark, R. Menikoff, and D. H. Sharp, Phys. Rev. {\bf D 22}
, 3012 (1980).

\bibitem{6.5}  C. Grosche, Phys. Rev. Lett. {\bf 71} 1 (1993).

\bibitem{7}  D. H. Lin, J. Phys. {\bf A {30}}, 4365 (1997).

\bibitem{8}  H. Kleinert, Phys. Lett. {\bf A 212}, 15 (1996).

\bibitem{9}  D. H. Lin, J. Phys. {\bf A 30}, 3201 (1997); J. Phys. {\bf A 31}%
, 4785 (1998); hep-th/9708144; hep-th/9709152.

\bibitem{10}  A. Inomata, H. Kuratsuji, and C. C. Gerry, ``{\it Path
Integrals and Coherent States of SU(2) and SU(1,1)}'', World Scientific,
Singapore, 1992.

\bibitem{11}  H. Kleinert, ``{\it Path Integrals in Quantum Mechanics,
Statistics and Polymer Physics}'', World Scientific, Singapore, 1995.

\bibitem{12}  H. Bateman, ``{\it Higher Transcendental functions}'',
McGraw-Hill, New York, 1953, Vol. II Ch. XI and N.H. Vilenkin, ``{\it %
Special Functions and the Theory of Group representations}'', Am. Math.
Soc., Providence, R I, 1968.

\bibitem{13}  W. Magnus, F. Oberhettinger, R.P. Soni, ``{\it Formuslas and
Theorems for the Special}

{\it \ Functions of Mathematical Physics}, Springer'', Berlin, 1966.

\bibitem{14}  L. S. Schulman, ``{\it Techniques and Applications of Path
Integrals}'', John Wiley sons, New York, 1981.
\end{thebibliography}
\end{document}